\documentclass[a4paper]{jpconf}
\usepackage{graphicx}
\begin{document}
\title{Equation of state and singularities in FLRW cosmological models}

\author{Leonardo Fern\'andez-Jambrina}

\address{ETSI Navales\\ Universidad Polit\'ecnica de Madrid\\ Arco de 
la Victoria s/n, 28040-Madrid, Spain}
\ead{leonardo.fernandez@upm.es}

\author{Ruth Lazkoz}

\address{F\'\i sica Te\'orica\\ Facultad de Ciencia y Tecnolog\'\i a\\ Universidad del
Pa\'\i s Vasco,\\ Apdo.  644, E-48080 Bilbao, Spain}

\ead{ruth.lazkoz@ehu.es}

\begin{abstract}
We consider FLRW cosmological models with standard Friedmann
equations, but leaving free the equation of state.  We assume that the
dark energy content of the universe is encoded in an equation of state
$p=f(\rho)$, which is expressed with most generality in the form of a
power expansion.  The inclusion of this expansion in Friedmann
equations allows us to construct a perturbative solution and to relate
the coefficients of the equation of state with the formation of
singularities of different types.\end{abstract}

\section{Introduction}
Since the end of the 20th century there has been an increase of
evidence for an accelerated expansion of the universe \cite{accel}. 
These puzzling observations do not fully fit in our present theoretical 
framework and therefore there are two trends of development 
attempting to ammend it.

One trend is grounded on modifications of general relativity as the 
correct theory of gravity \cite{modgrav}. The other major trend 
assumes the validity of general relativity and postulates the 
existence of an exotic component in the content of the universe known 
as dark energy \cite{darkfluid}.  

So far the only final state for our universe could be either infinite 
expansion or a  final collapse, the big crunch. But these new 
scenarios open the possibility of a different final fate, since both 
violate some of the classical energy conditions \cite{Nojiri:2005sx}. 
New candidates for a final fate of our universe are the big
rip \cite{Caldwell:2003vq}, sudden  singularities\cite{sudden},
big brake \cite{brake}, big freeze \cite{freeze}, inaccesible
singularities \cite{mcinnes}, directional singularities \cite{hidden},
$w$-singularities \cite{wsing}, in braneworld
models \cite{brane}, among others.

In \cite{modigravi} we dealt with such singular scenarios from the
point of view of geodesic completeness in modified theories of
gravity.  In this talk we do it following the approach of a dark
energy fluid.  More details can be found in \cite{darkeq}.

\section{Equations of state}
Instead of restricting to a specific model, we propose a 
framework which may comprise most of them.
With this aim in mind, we assume the validity of standard Friedmann 
equations,
\begin{equation}
H^2=\rho,\qquad \dot \rho+3H(\rho+p)=0,\end{equation}
assuming that pressure $p$ has a power
series in density $\rho$ of the matter content around a value $\rho_{*}$, for which a qualitative 
change of behaviour is expected,
\begin{equation}\label{eqstate}
p=p_{0}(\rho-\rho_{*})^{\alpha_{0}}+p_{1}(\rho-\rho_{*})^{\alpha_{1}}+\cdots,\end{equation}
where the exponents are real and ordered,
$\alpha_{0}<\alpha_{1}<\cdots$. 

Combining cosmological equations with our power expansion, we get
\begin{equation}\dot\rho=
-3\sqrt{\rho_{*}}\left(\rho_{*}+\frac{3}{2}(\rho-\rho_{*})+
\frac{3}{8}\frac{(\rho-\rho_{*})^2}{\rho_{*}}+\cdots
+p_{0}(\rho-\rho_{*})^{\alpha_{0}}+p_{1}(\rho-\rho_{*})^{\alpha_{1}}
+\cdots\right).\label{fried}
\end{equation}

We may integrate it for $\rho$ in 
powers of $(t_{0}-t)$, where $t_{0}$ satisfies 
$\rho_{*}=\rho(t_{0})$, if it exists,
\[
\rho(t)=\rho_{*}+\rho_{1}(t_{0}-t)^{\beta_{1}}+\rho_{2}(t_{0}-t)^{\beta_{2}}+
\cdots.
\]
In order to compare our results with previous ones in terms of 
expansions of the 
scale factor of the universe \cite{visser},
\begin{equation}
a(t)=a(t)=c_{0}|t-t_{0}|^{\eta_{0}}+c_{1}|t-t_{1}|^{\eta_{1}}+\cdots,
\end{equation}
we are to relate both series through Friedmann equation,
\begin{eqnarray}\nonumber
\rho=\frac{\eta_{0}^{^2}}{(t-t_{0})^2}+
\frac{2c_{1}}{c_{0}}\eta_{0}(\eta_{1}-\eta_{0})(t_{0}-t)^{\eta_{1}-\eta_{0}-2}+
\frac{2c_{2}}{c_{0}}\eta_{0}(\eta_{2}-\eta_{0})(t_{0}-t)^{\eta_{2}-\eta_{0}-2}+\cdots,
\end{eqnarray}

These expansions cope with almost every cosmological model, except 
those with oscillatory behaviour as the ones in \cite{wands}.
The results are consigned in table \ref{translate}.
\begin{table}[h]
\caption{\label{translate}Expansions of density and scale factor.}
\begin{center}
   \begin{tabular}{cccccc}
   \br
   ${\rho_{*}}$ & ${\beta_{1}}$ & $\beta_{2}$ & $\eta_{0}$ &
   $\eta_{1}$ & $\eta_{2}$ \\
   \mr
   0 & $(-\infty,-2)$ & $(\beta_{1},\infty)$ &
   - & - & -\\ 
   0 & $-2$ & $(-2,\infty)$ &
   $\pm \sqrt{\rho_{1}}$ & $(\eta_{0},\infty)$ & $(\eta_{1},\infty)$\\ 
   $0$ & $(-2,0)\cup(0,\infty)$ &   $(\beta_{1},\infty)$ &   0 & 
   $\frac{\beta_{1}+2}{2}$ & $(\eta_{1},\infty)$ \\
   $\neq0$   & $(0,1)$ & $(\beta_1,\infty)$ & 0 & 1 & $(1,2)$ \\
   $\neq0$   & 1 & $(1,\infty)$ & 0 & 1 & $[2,\infty)$\\
   $\neq0$   & $(1,\infty)$ &  $(\beta_{1},\infty)$ & 0 & 1 & 2 \\\br
   \end{tabular}
\end{center}
\end{table}

\section{Classification of singularities}
We follow the popular classification of singularities in 
\cite{Nojiri:2005sx}:
\begin{itemize}    
   \item Big bang / crunch: zero $a$, divergent $H$, density and 
   pressure.
   
   \item  Type I: ``Big rip'': divergent $a$. \cite{Caldwell:2003vq}

   \item  Type II: ``Sudden'': finite $a$, $H$, density, divergent 
   $\dot H$ and pressure. \cite{sudden}

   \item  Type III: ``Big freeze'': finite $a$, divergent $H$, 
   density and pressure. \cite{freeze}

   \item Type IV: ``Big brake'': finite $a$, $H$, $\dot H$, density, 
   pressure, divergent higher derivatives. \cite{brake}
\end{itemize}

In \cite{puiseux} a detailed analysis of the strength of these 
singularities is performed, making use of the concepts of geodesic 
completeness \cite{HE} and the definitions of strong singularities due 
to Ellis, Schmidt, Tipler and Kr\'olak \cite{strong}. Sudden and big brake singularities cannot be considered as such 
\cite{weaksudden}, since the cosmological spacetime may be continued 
beyond the singular event. And particles travelling at the speed of 
light do not experience the big rip, since they need infinite proper 
time to reach it \cite{puiseux}. These results may be checked in 
table \ref{not} in terms of the exponents of the  expansion of 
the density. Conformal diagrams for these 
models can be found in \cite{scott}.
\begin{table}[h]
\caption{\label{not}Singularities and density expansions.}
\begin{center}
   \begin{tabular}{cccccc}
   \br
   ${\rho_{*}}$ & ${\beta_{1}}$ & \textbf{Tipler} &
   \textbf{Kr\'olak} & \textbf{N.O.T.} \\
   \mr
   No & $(-\infty,-2)$ & 
   Strong & Strong & Big crunch/rip\\ 
   No & $(-2,0)$ &      Weak & 
   Strong  & III \\
   No & $(0,2)$ &      Weak & 
    Weak & II \\
   No & $[2,\infty)$ &      Weak & 
    Weak & IV \\
    Yes  & $(0,1)$ &  Weak & Weak & II \\
    Yes  & $[1,\infty)$ &  Weak & Weak & IV\\\br
   \end{tabular}\end{center}
\end{table}

\section{Singularities and equations of state}
Now we have all the ingredients for checking the formation of 
singularities in cosmological models with an equation of state of the 
form (\ref{eqstate}). We may integrate (\ref{fried}) in terms of 
power series of time and get the relevant exponents and coefficients. 
Comparing them with the results in table \ref{not}, we are ready to 
obtain the types of singularities which may appear in these models. 
The results may be found in table \ref{model}.
\begin{table}[h]
\caption{Singularities and equation of state.\label{model}}
\begin{center}
   \begin{tabular}{ccccc}
   \br
   $\rho_{*}$ & $p_{0}$ &${\alpha_{0}}$ & $\beta_{1}$  &\textbf{N.O.T.} \\
   \mr
   $0$ & Any & $(-\infty,0)$ &  
   $2/(1-2\alpha_{0})$ &  II\\ 
   $0$ & Any & $[0,1/2)$ &  
   $2/(1-2\alpha_{0})$ & IV\\ 
   Yes & Any & $(-\infty,0)$ &   
   $(1-\alpha_{0})^{-1}$   & II\\ 
   Yes & $\neq-\rho_{*}$ & $0$ &  
   $1$   & IV\\ 
    Yes & $-\rho_{*}$ & $0$ &   $1/(1-\alpha_{1})$ & IV \\ 
   Yes & Any & $(0,\infty)$ &  $1$  & IV\\ 
   \br
   \end{tabular}\end{center}
\end{table}
\section{Conclusions}
In this talk a detailed classification of the future behaviour of FLRW
cosmologies in terms of the equation of state is provided.  It depends
at most on the exponents of the first two terms of a power expansion
of the equation of state.  The scheme provides an easy route to
conclude the sort of singular behaviour.

\section*{Acknowledgments}
R.L. is supported by the University of the Basque Country through
research grant GIU06/37 and by the Spanish Ministry of Education and
Culture through research grant FIS2007-61800.

\end{document}